\documentclass{caosp305}

\usepackage{graphicx}
\usepackage{natbib}
\usepackage{caosp}

\articleNo{1}
\pubyear{2018}
\volume{48}
\volnumber{1}
\firstpage{1}
\received{22 October, 2017}
\accepted{23 October, 2017}

\def\BibTeX{{\rm B\kern-.05em{\sc i\kern-.025em b}\kern-.08em
             T\kern-.1667em\lower.7ex\hbox{E}\kern-.125emX}}

\begin{document}

%
\hauthor{"Stars with a stable magnetic field: from PMS to compact remnants", Brno, August 2017}

\title{Multiple, short-lived ``stellar prominences'' on the O giant $\xi\,$Persei: a magnetic star?}


\author{
        N.\,Sudnik\inst{1}
      \and 
        H.\,F.\,Henrichs \inst{2}   
       }

%
\institute{
           Belarussian State Pedagogical University\\
           220050, Sovetskaya 18, Minsk, Belarus \email{snata.astro@gmail.com}
         \and 
           Anton Pannekoek Institute for Astronomy, University\,of Amsterdam,
           Science Park 904, 1098 XH Amsterdam, Netherlands \email{h.f.henrichs@uva.nl}
          }

\date{October 20	, 2017}

\maketitle

\begin{abstract}
We present strong evidence for a rotation period of 2.0406 $\,$d of the O giant $\xi\,$Persei, derived from the N\,IV $\lambda$1718 wind line in 12 yr of IUE data. We predict that $\xi\,$Per has a magnetic dipole field, with superposed variable magnetic prominences. Favorable dates for future magnetic measurements can be predicted. We also analysed time-resolved He\,II 4686 spectra from a campaign in 1989 by using the same simplified model as before for $\lambda\,$Cephei, in terms of multiple spherical blobs attached to the surface, called stellar prominences \citep{SH2016}. These represent transient multiple magnetic loops on the surface, for which we find lifetimes of mostly less than 5\,h.  
\keywords{stars: early-type -- stars: individual: $\xi\,$Persei -- stars: magnetic field -- stars: winds, outflows -- stars: rotation}
\end{abstract}

\section{Stellar rotation and wind variability}
\begin{figure}
$
\begin{array}{cc}
    \includegraphics[width=0.5\textwidth,clip=]{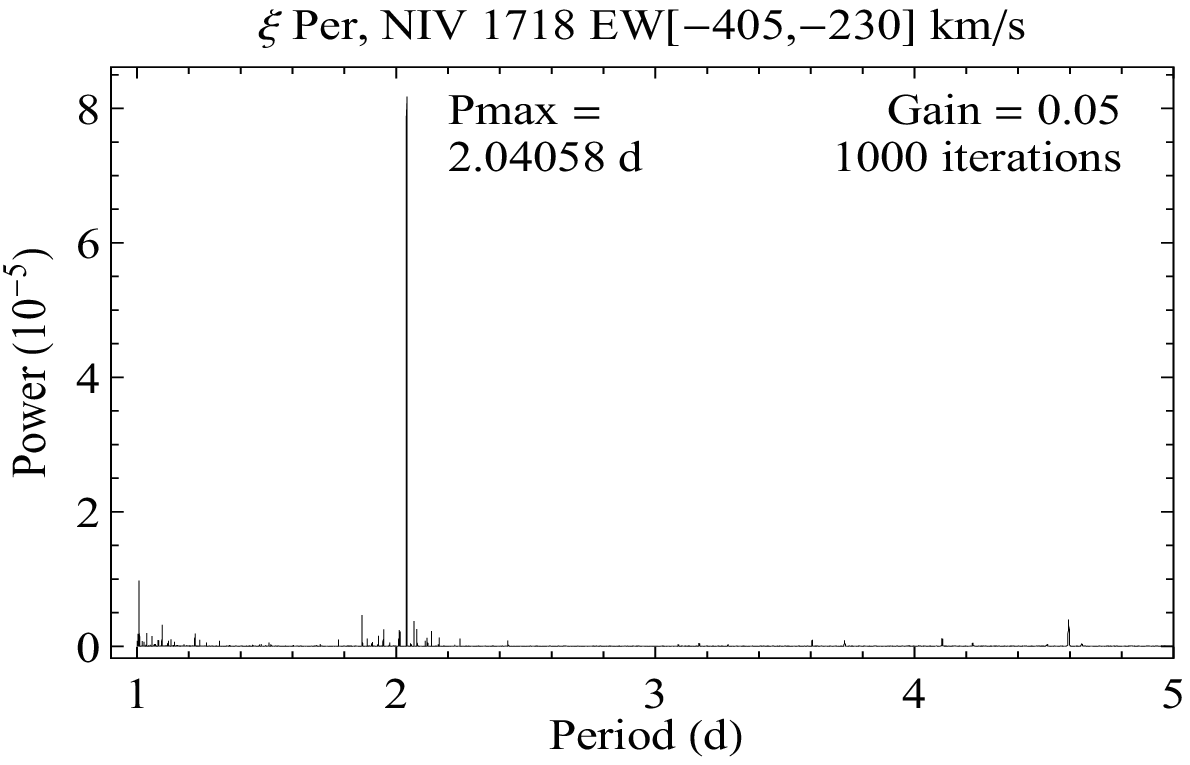}\\[-11em] 
    \includegraphics[width=0.5\textwidth,clip=]{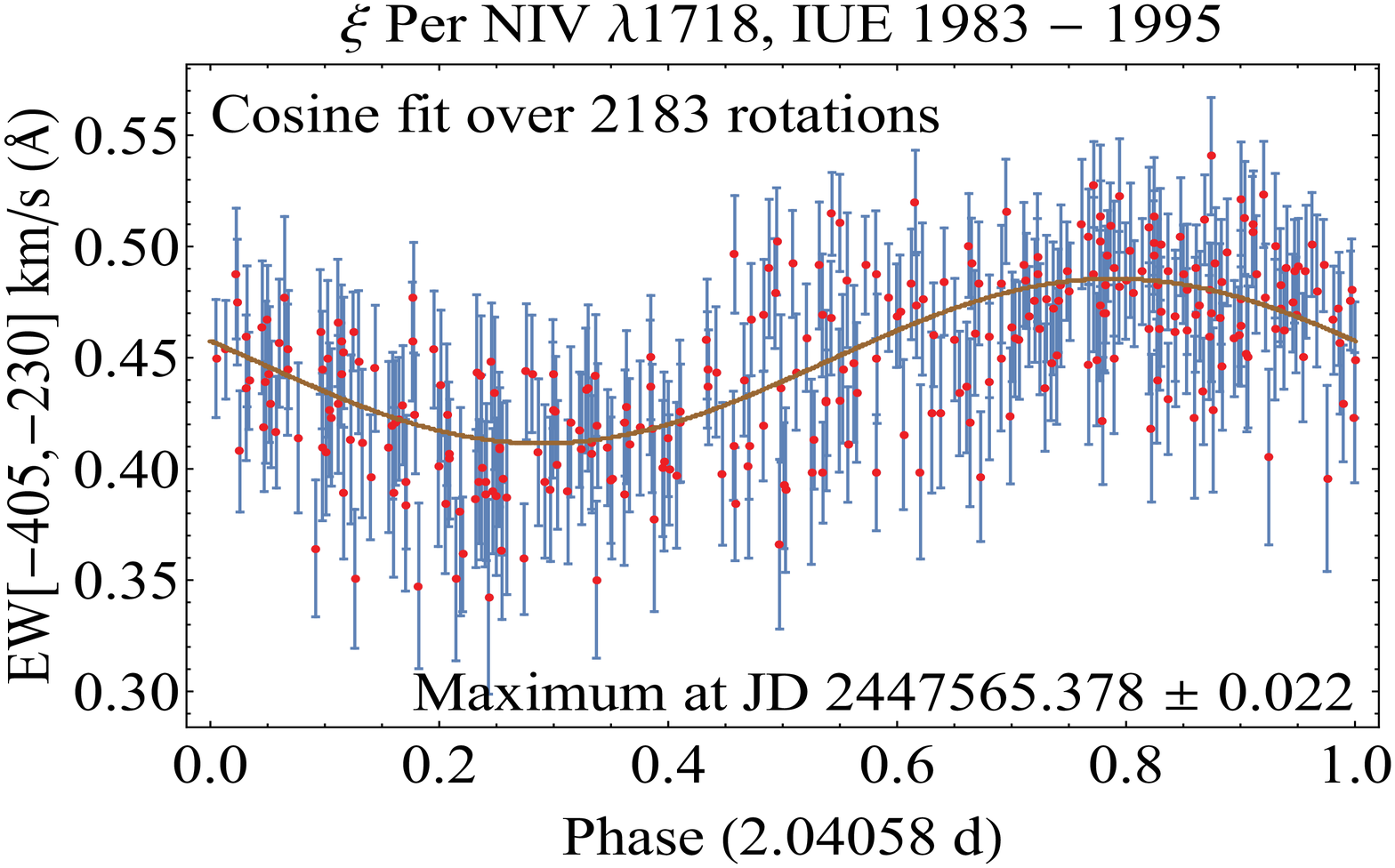}          
 &
    \includegraphics[width=0.5\textwidth,clip=]{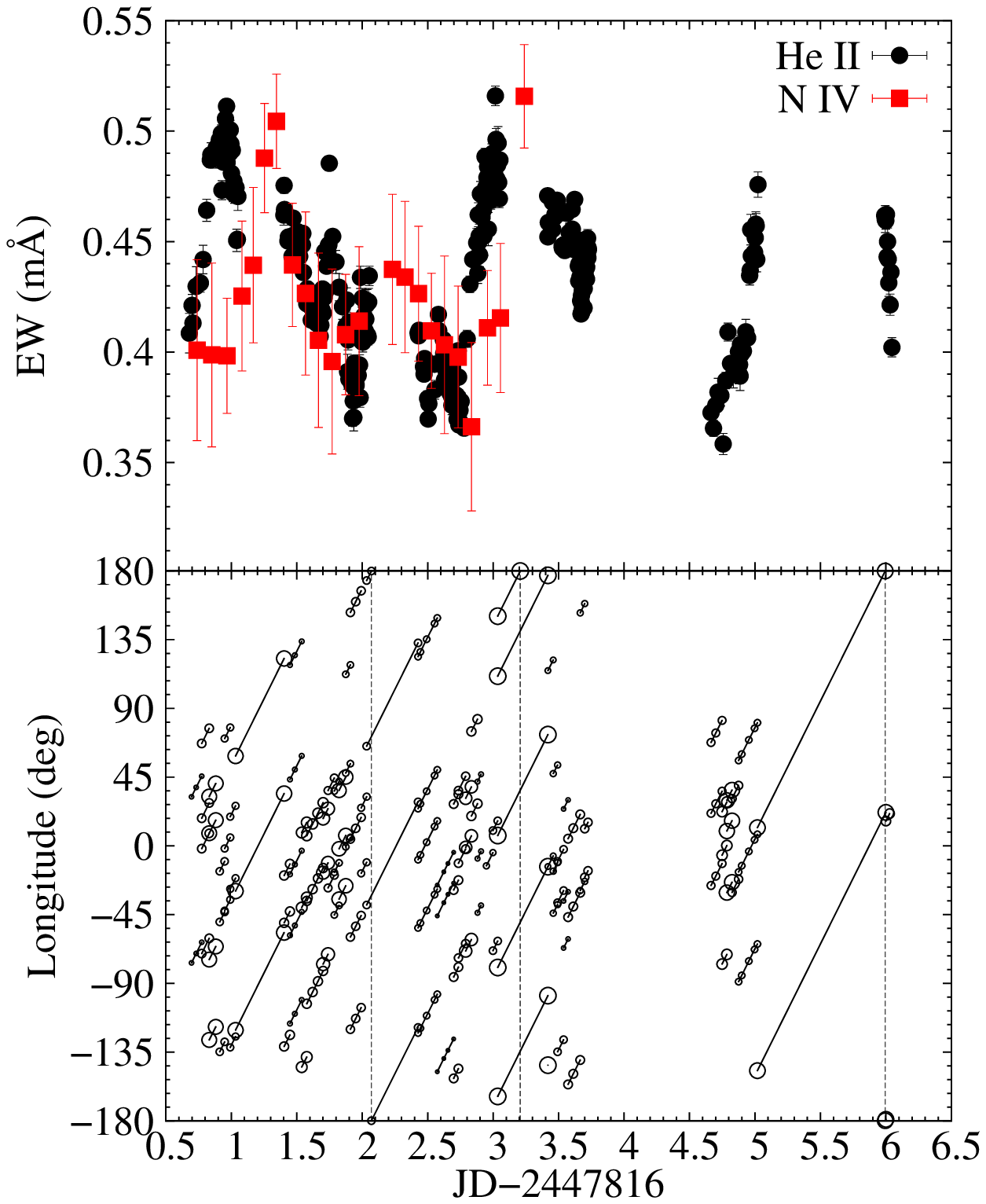}          
\end{array}
$           
\caption{{\it Top left:} Cleaned power spectrum of N\,IV $\lambda$1718 EWs. {\it Bottom left:} Phase plot of 307 datapoints over 12 years. {\it Top right:} Overplot of scaled N\,IV and He\,II EWs in 1989, showing a similar trend. {\it Bottom right:} Model fit results of subsequent quotient He II spectra displayed as circles at the fitted stellar longitude with 0$^{\circ}$ at the line of sight and size proportional to the fitted optical depth ($0.08 < \tau < 0.38$).	}
\label{fig}
\end{figure}

The well documented cyclic variability on the estimated rotational timescale in UV wind lines of early-type stars still lack a proper explanation. Except for the $\sim7\%$ of O stars with a dipolar magnetic field \citep{grunhut2017}, the unknown rotation period often hinders modeling of the many variable surface phenomena, which likely drive the wind variability. To search for the rotation period of the O7.5III(n)((f)) star $\xi$ Per ($v$sin$i = 230\,$km\,s$^{-1}$) we analysed the equivalent-width (EW) variations of the N\,IV $\lambda$1718 line in 307 IUE spectra over 12 years in the velocity range [$-$405\,km\,s$^{-1}$,$-v$sin$i$], which represents the lower wind. In the power spectrum (Fig.\,\ref{fig}, top left), we identify the peak at $2.04058\,$d as the rotation period. We can exclude twice this value, as often used before, because of the constraining stellar radius of $\sim$11 R$_\odot$. The phase-folded data shows a sinusoidal behavior (Fig.\,\ref{fig}, bottom left, which includes the ephemeris). The most likely explanation is that the star hosts a magnetic field with maximum strength at maximum EW. For a dipolar field, the footpoints of strong UV DACs should stem from only one of the magnetic poles. With the implied $i = 56^{\circ}$, and only one pole visible, $\beta$ should be near $\sim 90^{\circ} - i = 34^{\circ}$. Such a field has been hitherto undetected, possibly because of its weakness, and/or because all magnetic measurements so far have been taken at unfavorable phases.

\section{Model fits of He\,II {\boldmath $\lambda$}4686 spectra}

We applied the same stellar prominence model as for $\lambda\,$Cep \citep{SH2016} to 322 He\,II $\lambda$4686 spectra with 6 d coverage in 1989 of  $\xi\,$Per. To fit subsequent quotient spectra, multiple ($\leq$ 5) prominences with lifetimes up to 5\,h are needed (Fig.\,\ref{fig}, bottom right). These  are proposed to be at the footpoints of weaker intermediate DACs. Cancellation effects may make magnetic detection of these prominences difficult \citep{KS2013}. Similar behavior is also observed in other O stars, which suggests a common phenomenon.

\bibliography{Sudnik}

\begin{thebibliography}{3}
\expandafter\ifx\csname natexlab\endcsname\relax\def\natexlab#1{#1}\fi

\bibitem[{{Grunhut} {et~al.}(2017){Grunhut}, {Wade}, {Neiner}, {Oksala},
  {Petit}, {Alecian}, {Bohlender}, {Bouret}, {Henrichs}, {Hussain},
  {Kochukhov}, \& {MiMeS Collaboration}}]{grunhut2017}
{Grunhut}, J.~H., {Wade}, G.~A., {Neiner}, C., {et~al.} 2017, {\it \mnras},
  {\bf 465}, 2432

\bibitem[{{Kochukhov} \& {Sudnik}(2013)}]{KS2013}
{Kochukhov}, O. \& {Sudnik}, N. 2013, {\it \aap}, {\bf 554}, A93

\bibitem[{{Sudnik} \& {Henrichs}(2016)}]{SH2016}
{Sudnik}, N.~P. \& {Henrichs}, H.~F. 2016, {\it \aap}, {\bf 594}, A56

\end{thebibliography}
%

\end{document}